# Evaluating Attentional Impulsivity: A Biomechatronic Approach

Fateme Zare, Paniz Sedighi, and Mehdi Delrobaei

*Abstract*—Executive function, also known as executive control, is a multifaceted construct encompassing several cognitive abilities, including working memory, attention, impulse control, and cognitive flexibility. To accurately measure executive functioning skills, it is necessary to develop assessment tools and strategies that can quantify the behaviors associated with cognitive control. Impulsivity, a range of cognitive control deficits, is typically evaluated using conventional neuropsychological tests. However, this study proposes a biomechatronic approach to assess impulsivity as a behavioral construct, in line with traditional neuropsychological assessments. The study involved thirty-four healthy adults, who completed the Barratt Impulsiveness Scale (BIS-11) as an initial step. A low-cost biomechatronic system was developed, and an approach based on standard neuropsychological tests, including the trail-making test and serial subtraction-by-seven, was used to evaluate impulsivity. Three tests were conducted: WTMT-A (numbers only), WTMT-B (numbers and letters), and a dual-task of WTMT-A and serial subtraction-by-seven. The preliminary findings suggest that the proposed instrument and experiments successfully generated an attentional impulsivity score and differentiated between participants with high and low attentional impulsivity.

*Index Terms*—Biomechatronic systems, executive functions, cognitive assessment, attentional impulsivity, behavioral task measures, RFID tracking.

## I. INTRODUCTION

EXECUTIVE FUNCTIONS refer to a group of cognitive processes are essential for higher-order mental operations. The prefrontal cortex plays a vital role in developing executive function [1]. These cognitive processes include working memory, attention, self-control, cognitive flexibility, and are critical predictors of multitasking behavior [2]. Activities of daily living such as navigation, handling emotions, and learning require executive functioning.

Impaired executive function can lead to an inability to multitask, difficulty with attention, and socially inappropriate behavior. Lesions in the prefrontal cortex are linked to memory weakness, impulsiveness, attention deficits, and poor planning, highlighting the importance of the prefrontal cortex in the development of executive functions [3]. Furthermore, executive function impairments are critical for psychological science as they are related to most forms of psychopathology [4]. Impulsiveness is a symptom of various disorders such as attention deficit hyperactivity disorder (ADHD), borderline personality disorder (BDP), antisocial personality disorder, and bipolar disorder (BD), and is generally characterized as a preference to act without thinking, indicating poor self-control of behavior and leading to immediate reactions regardless of consequences [6], [7]. Impulse control disorders are also identified as one of the psychological symptoms of dementia (BPSD) [8].

Self-administered questionnaires such as the Barratt Impulsiveness Scale (BIS-11) [9], the UPPS Impulsive Behavior Scale [10], or the Eysenck Personality Questionnaire (EPQ) [11] can be used to assess impulsiveness. However, they are subject to inherent limitations, including generalisability, cultural impact, and unreliability of reports while expressing personality.

A variety of neuropsychological tests have been designed to examine executive functioning. However, psychometric tools in clinics and laboratories exhibit some limitations and dissimilarities between the observed performance of individuals in examinations and the actual performance in their daily life [14]. One of the limitations relates to their low ecological validity, which means a set of neuropsychological tests is required to predict the capacity of executive function in real life closely [15]. The evaluation of memory and attentional processes (e.g., selective, divided, and sustained attention) have been known as predictors of general performance in everyday life and require ecologically valid tasks [16].

Dual-task paradigms encompass a broad range of motor and cognition skills. The selection of a secondary task depends on the goal of the paradigm. Dual-task assessment captures the interaction of several aspects of executive functions and thus reaches the daily life requirements. The dual-tasking assessment method is crucial and has ecological validity [17]. Moreover, studying the motor-cognitive interface through motor and cognitive dual-task (MCDT) protocols enables clinicians to combine different motor and cognitive tasks and generate customized protocols [23].

There is limited research investigating the association between executive functioning, working memory, and impulsivity through MCDT protocols. This fact motivated us to develop a simple instrument and a set of cognitive-motor tasks to assess impulsivity. These tasks were inspired by standard neuropsychological tests such as Trail Making Test (TMT) and Serial Subtraction Test (SST). The main objective of this research is to assess executive function deficits, especially attentional impulsivity, using in-lab behavioral test scenarios based on standard neuropsychological tests. In this study, we:
- develop an accessible instrument and straightforward

*Corresponding Author: Mehdi Delrobaei.*
F. Zare is with the Faculty of Electrical Engineering, K. N. Toosi University of Technology, Tehran, Iran (e-mail:fatemezare79@email.kntu.ac.ir). P. Sedighi is with the Faculty of Electrical and Computer Engineering, University of Alberta, Alberta, Canada (e-mail: Sedighi1@ualberta.ca). M. Delrobaei is with the Mechatronics Department, Faculty of Electrical Engineering, K. N. Toosi University of Technology, Tehran, Iran, and the Department of Electrical and Computer Engineering, Western University, London, ON, Canada (e-mail: delrobaei@kntu.ac.ir). This work was supported in part by Iran National Science Foundation (INSF) under Grant 96S52820.

- cognitive-motor tasks based on standard neuropsychological tests;
- investigate the association between attentional impulsivity and task performance measures;
- explore the relationship between executive function and impulsivity.

The rest of this paper is organized as follows. Section II presents the related work. Section III describes our materials and methods. Section IV illustrates the experimental results. Section V validates the hypotheses and demonstrates limitations. Finally, the paper is concluded in Section VI, where future work opportunities are presented.

## II. RELATED WORK

Researchers have shown that traditional assessment methods may not be sufficient to accurately evaluate health components, including executive function, cognition, motor function, and task planning [19]. This section summarizes the benefits and challenges of the three fundamental assessment techniques in a structured literature survey.

### A. Physical-based assessment

This method assesses individuals by solving real-world problems in the lab or clinic settings. Perrochon *et al.* [20] aimed to determine if considering a challenging walking trial based on the Walking Trail-Making Test (WTMT) can be a potential detection instrument for cognitive impairment. They used the eight-meter electronic walkway to record the WTMT parameters and a webcam synchronized with the electrical walkway for error analysis. They showed that WTMT could provide early detection of cognitive impairment.

Klotzbier *et al.* [21] evaluated the Trail-Walking Test (TWT) as a potential detection tool for mild cognitive impairment (MCI). A stopwatch timed the trials, and recorded the mistakes. As a result, TWT identified a reliable measure in people with MCI.

Another research by Klotzbier *et al.* [22] aimed to determine dual-task performance for a complex Change-of-Direction walking task in children with Down syndrome. One of the disadvantages of this method is that clinicians may negatively affect the results. It is crucial to design technologies that allow patients to use them at home without the physical presence of the clinician. As a result, new technologies should be developed to enrich real environments [23]. Furthermore, the measurement frequency could be increased [24].

### B. Computer-based assessment

Technology generally enhances the effectiveness of traditional assessments. Additional digital features can be extracted in this method to analyze the task performance. Furthermore, more patients can be evaluated by this method. Hagler *et al.* [25] designed a computer game that can be used to assess various cognitive processes and evaluate the results of the pencil and paper TMT.

In [26], a computer game-based dual-task treadmill walking was proposed for testing executive function in patients with Parkinson's disease. Fellows *et al.* [27] created a tablet-based version of the TMT called the dTMT. The test measured the completion time, errors, and several digital performance measures, including the average duration of pauses and lifts and the drawing rate between circles.

Although digital devices are widely available and enable researchers to measure different aspects of human behavior accurately, variability in the motor and cognitive demands of the same tests affects the interpretation of the tests. Another challenge could be the hardware and software variability between devices that affect stimulus measurement [28].

### C. Virtual reality-based assessment

Virtual reality (VR) technology provides 3D real-world situations to immerse the subject in a real-life environment. The VR assessment method may overcome traditional measurements' lack of ecological validity.

Martelli *et al.* [29] created a VR floor maze test and analyzed the association between navigational skills and cognitive tests commonly used for executive functions. Parsons *et al.* [30] used the virtual reality Stroop task to examine differences in psychophysiological response. This platform showed that Stroop interference directly affects autonomic changes in psychophysiological arousal.

Plotnic *et al.* [31] a measure to study the cognitive-motor interactions by converting the standard version of TMT to a 3D VR-based format. S. de Leon-Martinez *et al.* [32] validated the Spheres & Shield Maze Task through VR to assess impulsivity and decision-making.

While VR technology appears to be an effective tool, several limitations, such as high implementation cost, design and development of software, and familiarization of participants with VR environments, may prevent researchers from adopting this technology widely [33].

Research has shown that dual-tasking assessment acceptance is increased if a test is short and delivered with simple instruments [34]. Over the last decade, experimental psychology studies have not fully covered the cognitive abilities involved in dual- and multitasking [35].

To our knowledge, the proposed approach in this work is the first study to combine the TMT and SST into a set of behavioral tasks to evaluate impulsivity. Furthermore, we used a low-cost RFID-based instrument to record data, making the measurements more accurate.

## III. MATERIALS AND METHODS

In this work, we employed a low-cost biomechatronic instrument and developed a set of behavioral tasks based on standard neuropsychological tests to evaluate impulsivity.

### A. Hypotheses

This work explicates three hypotheses:
1) The instrument and experiments proposed in this work can generate exclusive scores to evaluate attentional impulsiveness.

2) The proposed instrument and experiments can differentiate between the participants with high- and low-attentional impulsivity.
3) The proposed method in this work can associate the developed scores with some aspects of executive functioning skills.

### B. Questionnaire

Barratt Impulsiveness Scale (BIS-11) [9] was used as the gold standard to assess the personality construct of impulsiveness. The BIS-11 is a thirty-item self-report questionnaire with three sub scales of eight questions for attentional impulsivity, eleven for motor impulsivity, and eleven for non-planning impulsivity.

*Attentional impulsiveness* is the inability to concentrate or focus attention; *motor impulsiveness* is defined as acting uncontrollably, and *non-planning impulsiveness* is characterized as the lack of forethought [5].

The BIS-11 is rated on a four-point Likert scale of:
- never (1)
- occasionally (2)
- often (3)
- always (4)

### C. Participants

Seventy-seven university students initially filled out an online questionnaire. All the subjects were healthy, with no neurological conditions. Thirty-two healthy adults then volunteered to participate in the tests. The participants were not informed of the online questionnaire results before the tests. Ages ranged from 21 to 33 ($M = 24.17$; $SD = 2.98$), with 41% ($M = 23.4$; $SD = 3.68$) specified as female and 59% ($M = 24.70$; $SD = 5.01$) as male.

The tests were conducted for three weeks in the Biomechatronics Laboratory at K. N. Toosi University of Technology. Before instructing the participants to perform each task, they were asked to put on the designed instrument (an RFID-based glove and shoe system).

### D. System Architecture

In our previous work, a glove was designed as an object recognition tool, and the shoe as a navigation system to help the visually impaired [36], [37]. We extended our tool by integrating these components in this study and developed a new RFID-enabled wearable system. The system's general architecture is shown in Fig. 1.

The system consists of five main parts: two RFID transponders interacting with passive RFID tags, two processing units (Raspberry Pi), and the power source.

*1) RFID transponder:* The passive RFID tags operate without any built-in power sources and rely on the electromagnetic waves emitted by the transponder. Evaluation of the stability and sampling rate of a standard RFID transponder (MFRC522 unit) has shown that a signal could be captured with no information loss if:

$$F_s \leq \frac{f_{read}}{2} \quad (1)$$

where $F_s$ is the signal's frequency and $f_{read}$ is the read rate of the RFID transponder [38].

Our study used ISO 14443A standard 13.56 MHz passive tags and a 13.56 MHz MF-RC522 RFID transponder. The RFID transponder communicates with the processor via the serial peripheral interface (SPI) protocol.

*2) Processing units:* We selected Raspberry Pi 3B+ and 4B with 1.4 GHz and 1.5 GHz 64-bit quad-core processors. The employed processors showed comparable power consumption. The processors rely on removable micro SD cards, suitable for making customized onboard databases. The processors were also equipped with wireless networking onboard, which allowed us to locate the detected numbers and letters (pre-located on the passive RFID tags) on the web server.

*3) Power source:* The system's power consumption was calculated by considering the power consumption of the processors at idle and active modes, as well as the energy consumption of the RFID transponder while reading a tag. Raspberry 4B consumes 540 mA (2.7 W) at idle and 1.2 mA (6.4 W) at active mode for 400% CPU load. The current draw for Raspberry 3B+ is 350 mA (1.9 W) at idle and 1 mA (5 W) at active mode. Meanwhile, the energy an RFID module consumes is over 32 mA while reading passive tags. Given the average of 40% of the active mode (12W), the overall daily energy consumption is about 181.5 Wh. The system could last 3-4 hours with a 10000 mAh power supply.

The whole system was customized so the participants could conveniently perform the tasks. The instrument automatically recorded data during the experiments. The system elements can be seen in Fig. 2.

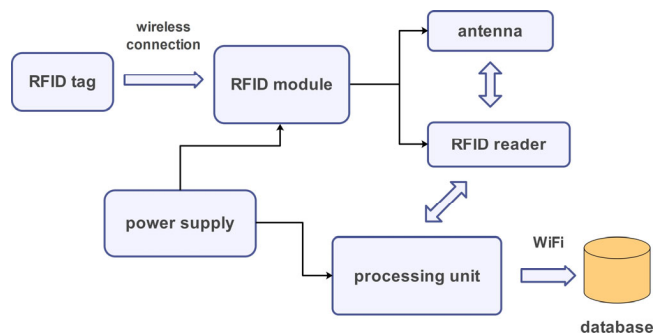

Fig. 1: The architecture of the employed RFID instrument; The surface of an RFID module comprises an antenna and an RFID reader. The antenna and reader gather and decode the tag information, then passed to the processor. The user can publish and save all the recorded data on a web server.

### E. Experimental Setup

Fig. 3 and Fig. 4 illustrate the experimental setups. The WTMT was designed in two parts: WTMT-A and WTMT-B, similar to the traditional TMT. In part A, we only placed the number labels from 1 to 20 on the floor. In part B, we randomly placed number labels from 1 to 20 and letter labels from A to T on the floor. The number and letter labels were randomly placed on a 35 $m^2$ area.

The third setup was designed based on the subtraction-by-seven (Serial-Seven Test). Fifteen number labels (27-90) were randomly placed on the table. The numbers and the letters were registered on the RFID passive tags mounted on the back of the labels. The participants' performances (order of numbers and letters and the time spent on each task) were accurately recorded and saved on a web server.

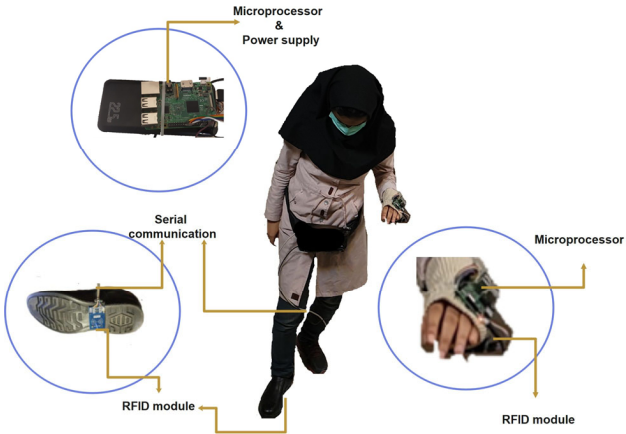

Fig. 2: The placement of the modules on the glove and the shoe; the processing unit (Raspberry Pi 3B) was located on the back, and the transponder was on the front of the glove. The battery and the second processing unit (Raspberry Pi 4B) were placed in a belt bag, and the transponder was placed on the bottom of the shoe.

*F. Experimental Protocol*

In Test 1 (WTMT-A), the participants were asked to step on the number labels in ascending order. In Test 2 (WTMT-B), the participants were instructed to step on the number and letter labels in ascending order alternatively Fig. 3. Furthermore, they were instructed to move from one target to the next as quickly and correctly as possible.

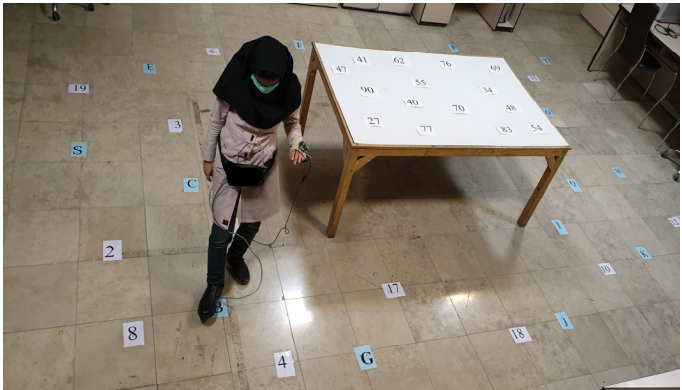

Fig. 3: The Walking Trail-Making Test; the numbers and the letters were randomly located on the floor. Participants performed a walking trail-making test while wearing the designed instrument. They were instructed to complete two parts. In part A, the participants were required to step on the labels in ascending order (1 to 20). In part B, they were required to go from a number to a letter in ascending order (from 1 to A, then to 2, and so on).

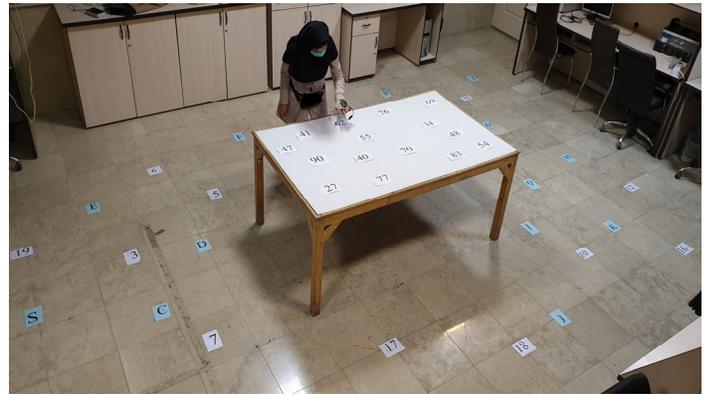

Fig. 4: Fifteen numbers ranging from 27 to 90 were placed on the table. Participants were instructed to step on target numbers on the floor (from one to ten), count by seven from ninety to twenty-seven reversely (serial subtraction-by-seven test), and touch the correct number while wearing the designed instrument. They were instructed to continue stepping on the next number on the floor and touching the next relevant number on the table.

In order to increase the attentional load, the third test was designed. The participants were instructed to do both the counting backward and WTMT-A tests simultaneously as follows:

- First, step on the target on the floor.
- Then count by seven from ninety to twenty-seven reversely and touch each number on the table with the glove.
- Continue this procedure until you touch ten targets on the table and step on ten labels on the floor (Fig. 4).

IV. RESULTS

Data were analyzed using the IBM SPSS Statistics software (version 25.0, IBM Inc., NY). We computed effect sizes for the independent t-test and the Mann-Whitney U test to analyze the magnitude of the experimental effect.

A large effect size shows the practical significance of the finding [39]. Statistical analyses were performed between two sets of data (the Barratt Impulsiveness Scale and the experimental results) to either accept or reject the hypotheses. Previous studies used latency and completion time differences between the paper-pencil TMT parts [40], [41]. To analyze the cognitive load impacts on the latency, we calculated the following:

1) T1, T2, T3: The completion time of each experiment
2) T4: Additional time for switching between numbers and letters in WTMT-B (the subtraction of completion time of WTMT-B (T2) and the time taken for connecting the numbers;
3) T5: Additional time for switching between numbers in WTMT-A and numbers in SST (the subtraction of completion time of WTMT-A and SST (T3) and the time taken for the numbers in WTMT-A;
4) E1: The number of errors that occurred in WTMT-B;
5) E2: The number of errors that occurred in WTMT-A and SST.

We assessed the normal distribution of parameters mentioned above using the Kolmogorov-Smirnov normality test.

The results indicated that only T1, T2, T3, and T5 were normally distributed.

The participants were initially divided into two groups with high and low attentional impulsivity scores. The BIS-11 score classification was based on the norm published in the work presented in [5].

The statistical confidence level was set to $p < 0.05$. The parametric correlation coefficient (Pearson - r) and non-parametric correlation coefficient (Spearman - rs) were used to evaluate the relationship between the two datasets.

The parametric correlation coefficient (Pearson - r) and non-parametric correlation coefficient (Spearman) were used to assess the relationship between the two datasets. The parameters E1, E2, and T4 were analyzed with a Mann-Whitney U test.

Table I illustrates no significant difference between genders since $p > 0.05$ and Cohen's d for the parameters show small effect sizes.

Table II reports the questionnaire subscale results. It presents the questionnaire results for each group, gender, and the total number of participants.

Table III shows the correlation analysis result. The correlation analysis shows that the parameters extracted from recorded data and the attentional impulsiveness subscale are associated.

Attentional impulsivity is significantly correlated with T2 (r = 0.67, $p < 0.01$), T3 (r = 0.78, $p < 0.01$), T5 (r = 0.656, $p < 0.01$) and moderately correlated with T1 (r = 0.408, $p < 0.05$). Also, the Spearman correlation coefficient for T4 was 0.629 ($p < 0.01$).

According to the preliminary correlation analysis, we defined an overall attentional impulsivity score and a task-switching score using parameters as:

Overall score:
$$S1 = \frac{T2}{N1 - E1} + \frac{T3}{N2 - E2} \quad (2)$$

Task-switching score:
$$S2 = \frac{T4}{N1 - E1} + \frac{T5}{N2 - E2} \quad (3)$$

where N1, N2 are the total numbers of labels in each test (forty numbers and letters in WTMT-B, and twenty numbers in the third experiment (WTMT-A and SST)).

The correlations analysis for the scores showed attentional impulsivity is significantly correlated with S1 (r = 0.787, $p < 0.01$) and S2 (r = 0.719, $p < 0.01$).

The independent t-test was used to compare the means of T1, T2, T3, T5 and the scores between the two groups. The mean differences in the t-test results between the two groups showed that the participants with higher attentional impulsiveness scores were generally more deficient in performing the tasks.

Table IV reports the results of the t-tests between the two groups. Levene's test for equality of error variances was performed. Results show a significant difference between the groups' mean in T2 ($t(18.46) = 5.39$, $p < 0.01$, Confidence Interval for Cohen's d = 1.069– 2.74), T3 ($t(30) = 8.03$, $p < 0.01$, Confidence Interval for Cohen's d = 1.857– 3.821), T5 ($t(22.54) = 6.36$, $p < 0.01$, Confidence Interval for Cohen's d

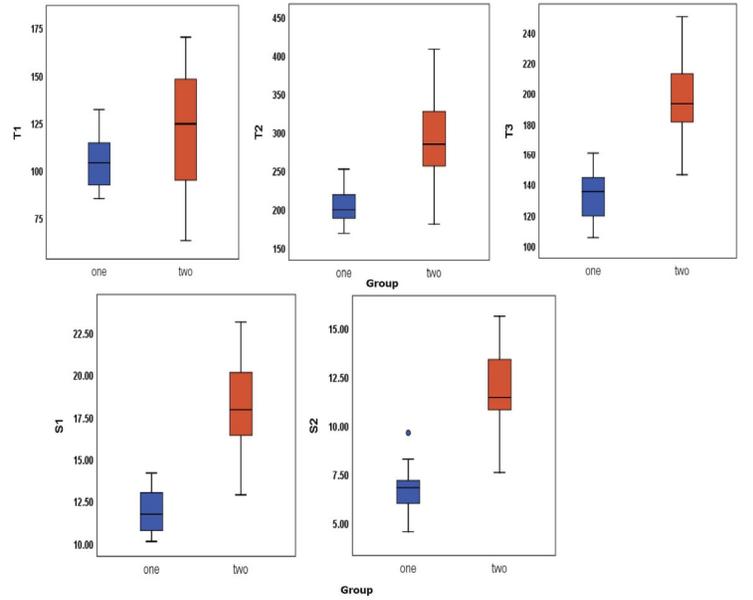

Fig. 5: The participants were divided into two groups based on high (group 1) and low (group 2) attentional impulsivity scores. Plots represent the differences in the mean parameters extracted from the experiments (T1, T2, T3, S1 and S2). T1, T2, and T3 are in seconds, whereas S1 and S2 indicate the overall and task switching score.

= 1.362−3.132), S1 ($t(21.307) = 7.94$, $p < 0.01$, Confidence Interval for Cohen's d = 1.832– 3.785), S2 ($t(23.315) = 7.8$, $p < 0.01$, Confidence Interval for Cohen's d = 1.788– 3.723) but no significant difference is observed between the mean of T1 between the groups ($t(20.538) = 1.78$, $p = 0.89$, Confidence Interval for Cohen's d = 0.078 ±.343).

Table V represents the Mann-Whitney U test results. A significant difference is observed between the mean rank of E1 (Cohen's d = 1.538, $\eta^2 = 0.372$, $p < 0.01$), E2 (Cohen's d = 0.747, $\eta^2 = 0.122$, $p < 0.05$) and T4 (Cohen's d = 2.219, $\eta^2 = 0.552$, $p < 0.01$) in the groups. The large effect sizes in the t-test and the Mann-Whitney U test results show a meaningful difference between groups. The large effect sizes indicate that our findings have practical significance.

Participants with higher attentional impulsivity scores show a greater mean than others with lower scores in the t-test and Mann-Whitney U test results. Fig. 5 compares the mean of the parameters of interest between the two groups.

## V. DISCUSSION

Generally, interactions of sensory and motor functions are involved in response to cognitive tasks; first, to choose the required behavioral reaction and then to arrange and execute them. As a result, evaluating such processes with traditional – paper-pencil tests or even computerized testing methods often falls short. Most everyday situations require different executive functions simultaneously or in rapid sequence.

The current study presents an ecological assessment of executive function in agreement with relevant neuropsychological tests in behavioral scenarios. The WTMT-A evaluates attention, visual scanning, and processing speed. In addition

TABLE I

GENDER DIFFERENCES ANALYSIS

|    | t-test Mean | t-test Std.Error | t-test t-value | Mann-Whitney U test Mean Rank | Mann-Whitney U test Sum of Ranks | Mann-Whitney U test U-value | effect size Cohen's d |
|----|------|----------|---------|-----------|--------------|---------|---------|
| T1 | 3.617 | 9.644 | 0.374 | | | | 0.13 |
| T2 | 36.85 | 23.85 | 1.547 | | | | 0.55 |
| T3 | 6.65 | 14.514 | 0.458 | | | | 0.16 |
| T5 | 6.4 | 12.829 | 0.5 | | | | 0.122 |
| S1 | 0.55 | 1.39 | 0.397 | | | | 0.14 |
| S2 | 0.646 | 1.12 | 0.577 | | | | 0.20 |
| T4 | | | | | | 72.5 | 0.13 |
| E1 | | | | 2.67 | 172 | 100 | 0.32 |
| E2 | | | | 6.4 | 228 | 98.57 | 0.34 |

T-test results for comparing T1, T2, T3, T5, S1, S2 and Mann-Whitney U test results for comparing E1, T4, and E2 between genders. E1 and E2 are the number of errors that occurred. T1, T2, T3, T4, and T5 are in seconds. No significant differences were observed.

TABLE II

QUESTIONNAIRE SUB-SCALES RESULTS

| subsacle | Impulsiveness Attentional | Impulsiveness Motor | Impulsiveness Non-planning |
|----------|------------|-------|-------------|
| Group 1 (n=16) | 15.562±1.67 | 18.375±1.962 | 20±3.812 |
| Group 2 (n=16) | 23.562±2.55 | 21.187±3.79 | 22.125±3.685 |
| Female (n=13) | 19.16±5.047 | 19.45±2.019 | 20.58±2.928 |
| Male (n=19) | 19.88±4.18 | 20±3.78 | 21.38±4.19 |
| Total | 19.59±4.61 | 19.78±3.29 | 21.06±3.84 |

Group 1 defines participants with low attentional impulsivity scores and Group 2 shows participants with high attentional impulsivity scores. Numbers are in $M \pm SD$ (M=mean, SD=standard deviation).

TABLE III

PEARSON CORRELATION COEFFICIENT ANALYSIS

|  | T1 | T2 | T3 | T5 |
|---|-----|------|------|------|
| Attentional Impulsiveness | **0.408*** | **0.67**** | **0.78**** | **0.656**** |
| Motor Impulsiveness | -0.08 | 0.32 | 0.279 | 0.193 |
| Non-planning Impulsiveness | 0.231 | 0.321 | 0.332 | 0.289 |
| T1 | 1 | **0.577**** | **0.536**** | 0.301 |
| T2 | | 1 | **0.802**** | **0.67**** |
| T3 | | | 1 | **0.903**** |
| T5 | | | | 1 |

Pearson correlation coefficient of questionnaire & parameters of interests. When the p-value is statistically significant, it is highlighted in bold ($*p < 0.05$, $**p < 0.01$, $***p < 0.001$).

to the assessed skills in part A, WTMT-B also evaluates cognitive flexibility and working memory [21]. The WTMT-B may also evaluate high-level visual processing and problem-solving [42]. The serial seven test is considered a measure of attention and working memory [43].

We aimed to develop a practical alternative to traditional tests in addition to investigating the following hypothesis.

*A. Hypothesis I*

The first hypothesis investigates if the proposed method can generate exclusive scores to evaluate attentional impulsivity. According to the findings, both the task-switching and the overall scores are highly correlated with the results of the attentional impulsiveness subscore of the BIS-11. Therefore, the first hypothesis of this research can be confirmed.

Previous studies have shown that impulsivity increases as the task's complexity increases, affecting attentional load during dual-task performance [44]. Our results support previous evidence. According to the correlation analysis results, the task parameters with a higher difficulty level are strongly associated with attentional impulsivity. We recommend employing the proposed scores consisting of the parameters of the WTMT-B and WTMT-A+SST in applications requiring the quantification of attentional impulsivity.

*B. Hypothesis II*

We were able to develop scores and quantify attentional impulsivity; now, the following hypothesis is whether the proposed method and performance measures can differentiate between the subjects with high- and low- attentional impulsivity. We initially divided the participants into two groups based on their questionnaire results. Statistical analyses were then carried out to investigate the validity of the second hypothesis.

This study found that the participants with low impulsivity had statistically significantly lower overall and task-switching scores and the participants with high impulsivity had substantially higher scores.

Furthermore, it is noted that we observed no significant difference between the groups in the task with lower cognitive load (WTMT-A), while experiments with higher cognitive loads (WTMT-B and WTMT-A+SST) showed a more significant difference. As the results of analyzing the tasks with the increased cognitive load were highly correlated with higher attentional impulsiveness scores, we conclude that the difficulty level of the cognitive assessments is a significant factor in evaluating impulsiveness.

*C. Hypothesis III*

Previous studies emphasized the effect of attentional impulsivity in general executive function and dual-task performance [45], [46]. Therefore, performing two tasks simultaneously (a dual-task cognitive scenario) causes demands on attention [47]. Previous findings also emphasized attentional impulsivity as a critical predictor of executive functioning skills [40], [48].

Cognitive flexibility is an essential aspect of executive function that considers the ability to switch between different tasks.

TABLE IV

THE INDEPENDENT T-TEST RESULTS

| t-test | T1 | T2 | T3 | T5 | S1 | S2 |
|---|---|---|---|---|---|---|
| t-statistic | 1.78 | 5.39** | 8.03** | 6.36** | 7.94** | 7.8** |
| Mean difference | 15.94 | 92.06 | 63.8 | 51.75 | 6.11 | 4.88 |
| Std.Error difference | 8.92 | 17.09 | 7.95 | 8.91 | 0.77 | 0.63 |

T-test results for comparing T1, T2, T3, T5, S1, S2 between groups (high and low attentional impulsiveness scores). The mean and standard error differences of T1, T2, T3, and T5 are in seconds. Significant differences is denoted by *, ** and *** (* if $p < 0.05$, ** if $p < 0.01$ and*** if $p < 0.001$).

TABLE V

THE MANN-WHITNEY U TEST RESULTS

| Mann-Whitney U test | E1 | E2 | T4 |
|---|---|---|---|
| U-statistic | 36.5** | 75.5* | 16.5** |
| Mean Rank difference | 11.44 | 6.56 | 13.94 |
| Sum of Ranks difference | 163 | 105 | 223 |

The Mann-Whitney U test results for the comparison of E1, T4 and E2 between groups (High and low attentional impulsiveness scores). E1 and E2 are the number of the errors that occurred. T4 is in seconds. Significant differences is denoted by *, ** and *** (* if $p < 0.05$, ** if $p < 0.01$ and *** if $p < 0.001$).

We believe our third hypothesis could explore the possible association between impulsivity and cognitive flexibility. The correlation analysis demonstrated that attentional impulsivity and parameters related to task-switching ability (T4, T5, S2) are significantly correlated.

The mean differences in the independent t-test and the Mann-Whitney U test results indicated that the group with higher attentional impulsivity scores had weaker task-switching abilities.

Several cognitive processes, such as attention, switching ability and working memory, work together to implement cognitive flexibility [49]. Also, higher working memory capacity leads to better cognitive flexibility [50]. Therefore, our results provides some evidence that attentional impulsivity impacts working memory and attention. Our results indicate that executive functions and attentional impulsiveness were associated.

*D. Limitations*

Some limitations of the proposed work need further improvement: (1) lack of sensors to monitor gait characteristics (step velocity, step length, step time). We required more information regarding motor functions to prove the association between multitasking ability and impulsiveness. As a result, we could not exclusively accept or reject the third hypothesis. (2) Our small sample size might have made it difficult to determine if this study's outcomes are all valid. (3) The proposed approach was not validated as a neuropsychological assessment method in a clinical setting.

VI. CONCLUSION AND FUTURE WORK

This work examined the feasibility of using simple instruments and behavioral tasks to assess attentional impulsivity. The motivation for this work was mainly to employ accessible technologies and protocols to enrich clinical cognitive assessment methods. The key contributions of our work are as follows.

We developed a new cognitive-motor approach and impulsivity score inspired by traditional neuropsychological tests (TMT and SST). To the best of our knowledge, this is the first study that integrates these two tests to examine executive function deficits. Moreover, this study provided an engaging paradigm to evaluate attentional impulsivity in real-world scenarios.

We recommend using the WTMT with the higher cognitive load based on our sensitivity results to evaluate attentional impulsivity further. It seems to be an ecologically valid dual tasking with excellent diagnostic ability to differentiate between individuals with and without attentional impulsivity.

In our future work, we plan to extend our experimental setup and protocols to implement more standard neuropsychological tests in real scenarios and define more concrete threshold values for clinicians' use.

ACKNOWLEDGMENT

The authors appreciate the participation of volunteers in this study.